\documentclass[fleqn]{aa}  
\usepackage{supertabular,booktabs}  
\usepackage{graphicx}
\usepackage{xcolor}
\usepackage{hyperref}
\usepackage{txfonts}
\usepackage{calligra}
\usepackage{calrsfs}
\usepackage[T1]{fontenc}
\usepackage[mathcal]{eucal}
\usepackage{longtable}
\usepackage{multicol}

\usepackage[switch, modulo]{lineno}

\sfcode`A=1000 \sfcode`B=1000 \sfcode`C=1000 \sfcode`D=1000
\sfcode`E=1000 \sfcode`F=1000 \sfcode`G=1000 \sfcode`H=1000
\sfcode`I=1000 \sfcode`J=1000 \sfcode`K=1000 \sfcode`L=1000
\sfcode`M=1000 \sfcode`N=1000 \sfcode`O=1000 \sfcode`P=1000
\sfcode`Q=1000 \sfcode`R=1000 \sfcode`S=1000 \sfcode`T=1000
\sfcode`U=1000 \sfcode`V=1000 \sfcode`W=1000 \sfcode`X=1000
\sfcode`Y=1000 \sfcode`Z=1000

\newcommand{\Msun}{\hbox{M$_\sun$}}
 
\begin{document}

   \title{A tight relation between the distribution of globular clusters and dark matter in  AS1063.}
   \titlerunning{Relation between lensing and globular clusters}
   \author{J.M. Diego\inst{1}\fnmsep\thanks{jdiego@ifca.unican.es}
       \and C. Goolsby \inst{2}
       \and C.J. Conselice \inst{2}
       \and J.M. Palencia \inst{1}
    }      
   \institute{Instituto de F\'isica de Cantabria (CSIC-UC). Avda. Los Castros s/n. 39005 Santander, Spain 
        \and Jodrell Bank Centre for Astrophysics, Alan Turing Building, University of Manchester, Oxford Road, Manchester M13 9PL, UK 
          }
 \abstract{
 Based on deep high resolution JWST images of AS1063, and after a careful masking of artifacts, extended features in the cluster, and background galaxies (including known lensed ones), we have identified tens of thousands of unresolved point sources in the central region of the galaxy cluster. We extended the identification of these point sources up to 1.18 Mpc from the center of the cluster using data in the second module. Most of these sources are expected to be globular clusters orbiting in the deep potential well of the cluster, but also the surviving compact cores of satellite galaxies. We study the distribution of the globular clusters and compared it with the distribution of mass from a lens model derived from the same JWST data. We find a very tight correlation between the two distributions, but also some differences, including a more concentrated distribution for the globular clusters than for dark matter. We explored the possibility of using the distribution of globular clusters as a proxy for the lensing mass. We find that a simple smoothing kernel can transform  the discrete distribution of point sources into a continuous two-dimensional distribution that matches well the lensing convergence.  This suggests that globular clusters can be used as tracers of the dark matter distribution in other massive clusters where gravitational lensing constraints are scarce but globular clusters can be detected more easily, for instance in low redshift galaxy clusters. 
   }

   \maketitle
%
\section{Introduction}

Data from the \textit{James Webb} space telescope (JWST) allows us to study galaxy clusters and the background galaxies they magnify with unprecedented detail.
Much attention has been paid to the latter, where some of the best studied so far galaxies at high redshift are found magnified behind powerful gravitational lenses. This has enabled a wealth of studies on these very young galaxies, including also detailed spectroscopy that takes advantage of the magnification provided by the cluster lenses. 

But the new JWST images also reveal a never seen before (at these redshifts) wealth of unresolved sources in the galaxy clusters themselves \citep{Lee2022,Harris2023}. Most of these compact sources are believed to be globular clusters (GC) that are stripped away from their host galaxies and, thanks to their very compact nature, can survive the strong tidal forces that ripped apart their host galaxies. Some of these point sources are also expected to be compact cores of galaxies surrounding the central super massive black hole (SMBH) of each galaxy. These very compact cores can survive close encounters with other galaxies, or with the galaxy cluster central region. Since self-gravitating systems have negative specific heats, during an interaction the heat transfer between the cluster core and the satellite galaxy can result in the satellite core contracting \citep[gravothermal instability][]{Lynden-Bell1968,Lightman1978,Ernst2007}, which together with tidal shocks accelerate the core contraction in the satellite galaxies \citep{Gnedin1999}, making them more prone to survive strong tidal forces. The surviving satellite cores end up being very compact and appear as unresolved sources in JWST images. 
For simplicity we refer to the two types of point-like objects as GCs, but keeping in mind that their nature and origin can be quite different. Owing to their compactness, it is very unlikely a GC collides with another galaxy in the cluster or that two GCs collide with each other, so the interaction with the surrounding medium is mostly only gravitational. Hence GCs behave as collisionless particles, pretty much as dark matter (DM) particles are expected to behave. This expected similarity in how GCs and DM interact with their surrounding medium begs the question about the possibility of using GCs as tracers of the distribution of the invisible DM, that may outperform the suggested correlation between the intracluster-light and DM \cite{Montes2019,Manuwal2025,Yoo2025}, which has been recently suggested to be a biased tracer of the DM distribution \cite{Butler2025}. Results from simulations indicate that such tight relationship between GC and DM  should exist \citep{Cote2003,Harris2017,ReinaCampos2022,ReinaCampos2023}, and observations with JWST start to show this relationship with actual data \citep{Diego2023d,Diego2024a}.  Very large numbers of GCs have been observed in nearby galaxy clusters \citep{Peng2011,Durrell2014}, each with an estimated number of ${\rm O}(10^5)$ GCs. \\ 

A correlation between the number of GCs, $N_{\rm gc}$, and  the virial mass of the host halo, $M_{\rm vir}$, is well established for a wide range of halo masses. 
From the relation of \cite{Burkert2020},
\begin{equation}
M_{\rm vir} = 5\times10^9N_{\rm gc}\, \Msun, 
\end{equation}
we expect a massive cluster with mass $M_{\rm vir}=10^{15}\, \Msun$ to host a staggering $N_{\rm gc} \sim 2\times10^5$ GCs within the virial radius. This number corresponds to a mean projected number density of $\approx 1$ GC per $6\times6$ kpc$^2$ unit area assuming a virial radius of 1.5 Mpc, or in volumetric units this corresponds to one GC per $42^3$ kpc$^3$ volume, or a mean distance of 84 kpc between GCs. The gravitational acceleration that one massive $10^7\, \Msun$ GC imparts onto another GC at 84 kpc distance is $\approx 10^5$ times weaker than the acceleration from a  $M_{\rm vir}=10^{15}\, \Msun$ cluster at 1.5 Mpc, thus making the attraction between two GCs, separated by the mean distance above, negligible compared with the attraction imparted by the galaxy cluster as a whole. Together with the fact that GCs are very compact ($r<50$ pc), close encounters between GCs are expected to be extremely rare. The rarity of two body interactions between GCs distinguishes their dynamics from the well studied stars inside an evolved globular cluster, which tend to converge to an state of energy equipartition due to two-body relaxation  \citep{King1962,Spitzer1969,Meylan1997,Baumgardt2017}, but see also \cite{Trenti2013} for an alternative perspective. 
Due to the low probability of two-body relaxation and the relatively narrow mass function of GCs, equipartition is not expected in the system of GCs in galaxy clusters. Some level of dynamical friction is certainly expected as the more massive GCs. Specially the galactic cores remnants hosting a SMBH with masses exceeding $10^8\, \Msun$, will be slowed down more than the lighter GCs, but this has not been studied in detail in galaxy clusters, so the situation is unclear. \\


The massive galaxy clusters that host GCs are also powerful gravitational lenses. Through strong lensing it is possible to map the mass distribution in the inner regions of the lenses (where strong lensing takes place) and compare it with the distribution of GCs. Hence, the expected correlation between the distribution of DM and GCs can be tested in galaxy cluster lenses. For a galaxy cluster to act as a strong lens, the surface mass density needs to exceed a critical value that scales inversely proportional with the distance to the cluster. Hence, low redshift clusters (where GCs can be studied best) never become supercritical and can not produce strong lensing. One needs to go to higher redshifts ($z\gtrsim0.2$) in order to use strong lensing to map the mass in cluster cores. 

Early studies of GCs in galaxy cluster lenses  were carried out by HST data, on relatively low redshift clusters, such as Abell 1689 (z=0.183), where a similarity between the profiles of the total projected lensing mass (mostly DM) and GCs was already demonstrated \citep{AlamoMartinez2013}. Similar studies based on HST are limited due to sensitivity of the telescope to the infrared (where evolved GCs are brightest) and the scarcity of low redshift cluster lenses. In aged stellar populations, as the ones expected in GCs within galaxy clusters, the emission peaks at 1.6 $\mu$m in the rest-frame of the GC \citep{Sawicki2002,Muzzin2013}, where HST's ability to detect point sources is at its worst. Meanwhile, JWST's larger mirror and sensitivity in the infrared makes it the ideal telescope to study GCs in galaxy cluster lenses. The first attempt was already made with the very first scientific observation of JWST, the galaxy cluster lens SMACS0723 \citep{Lee2022,Faisst2022}, quickly followed by a comparison with its lensing mass \citep{Pascale2022,Mahler2023,Diego2023d}. Many studies of GCs with JWST followed \cite{Harris2023,Martis2024,DAbrusco2025,Harris2025}. Other observatories, such as Euclid, have recently followed the study of  GCs in nearby galaxy clusters, such as Fornax \citep{EuclidGC2025}. While Euclid performs shallower observations,  
it has a much larger field of view, which is better suited for nearby clusters. For instance, at the distance of the Fornax cluster, an object appears 9.86 magnitudes brighter than if the same object is at $z=0.35$ (ignoring color corrections), thus compensating for the reduced depth. 
At the redshift of Fornax, no strong lensing is expected but weak lensing mass maps can be obtained by Euclid thanks to its very large sky coverage. Studies of nearby clusters comparing the distribution of GCs and DM (from weak lensing) can be attempted in the near future.

Among the strong lensing clusters showing an abundance of GCs in JWST data, AS1063 at $z=0.348$   \citep[also known as RXC J2248.7-4431][]{Guzzo2009}, is a remarkable example given the depth of JWST data and relatively lower redshift (distance modulus 41.32). This cluster was selected as one of the clusters for the Hubble Frontier Fields (HFF) program \citep{Lotz2017} due to its large mass and existence of known gravitationally lensed galaxies \citep{Postman2012a,Balestra2013,Moona2014,Johnson2014,Richard2014,Zitrin2015}. As part of this program,  the central $\sim 10$ arcmin$^2$ region was observed in wavelengths ranging from 0.45 $\mu$m to 1.6 $\mu$m and to a depth of $\sim 28.5$ mag in the visible and IR bands. The area observed by the HFF program around AS1063 was later doubled thanks to the Beyond Ultra-deep Frontier Fields and Legacy Observations (BUFFALO) program \citep{Steinhardt2020}, although with shallower observations than in the HFF program. 
JWST data complements the high-quality Hubble Space Telescope (HST) data extending the wavelength coverage to $\approx 5 \mu$m and similar depths. \\
The high-quality data provided by the HFF program allowed the identification of hundreds of multiply lensed galaxies behind the six clusters HFF clusters. Many of the new lensed galaxies were spectroscopically confirmed from the ground, with the Multi Unit Spectroscopic Explorer \citep[MUSE, ][]{Bacon2010} playing a pivotal role. Improved lens models for this cluster were derived from the HFF data  \citep{Caminha2016,Diego2016,Vega-Ferrero2019,Chan2020,Granata2022} making this cluster one of the best studied gravitational lenses. \\

AS1063 was recently observed by JWST as part of the GLIMPSE program \citep{Atek2025}, with very long exposures making AS1063 an ideal object for both lensing and GC studies. GLIMPSE data has been used for a variety of studies from high redshift galaxies \citep{Kokorev2025a,Kokorev2025b,Korber2025,Chemerynska2025}, including very low metallicity ones  \citep[or Pop III galaxy candidates][]{Fujimoto2025}, to early SMBH \citep{Fei2025}. The new JWST data allows for novel studies that could not be attempted before. Precise lens models of massive galaxy clusters tracing the projected distribution of mass (mostly DM), can be compared with the distribution of GCs detected by JWST. GCs are expected to react to the gravitational potential in a very similar fashion as DM does, assuming DM is composed of collisionless particles. Using GCs as visible tracers of DM shall open new avenues to advance on the understanding of this mysterious substance. 

Based on previous HST data and the new JWST data, a new lens model was derived in \cite{Diego2026c} (or paper-I). In this work we used this lens model to compare with the distribution of GCs, which is the main focus of this paper.   Here we present the GCs from the same GLIMPSE data and compare the distribution of GCs with the lensing mass from paper-I.\\
\begin{figure*} 
  \includegraphics[width=\linewidth]{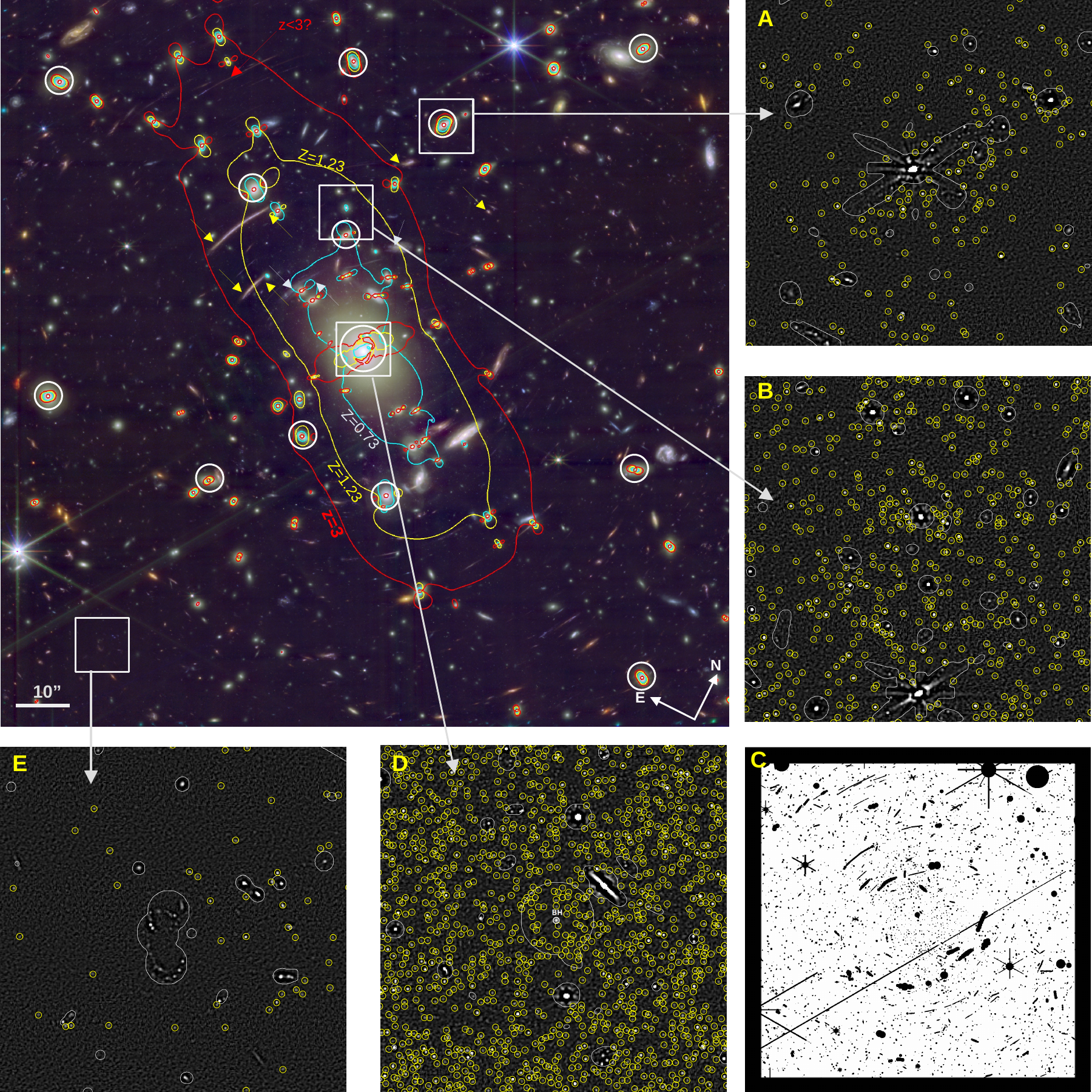} 
      \caption{Color image of AS1063 combining 12 HST and JWST filters (range 0.4--5 micron). Three critical curves from our lens model are shown at $z=0.73$ (cyan), $z=1.23$ (yellow), and $z=3$ (red). The yellow arrows mark two families of lensed galaxies at $z\approx1.23$ intersected by the critical curve at that redshift, while the cyan arrows mark one family of lensed images from a galaxy at $z=0.73$ intersected by the cyan critical curve. White circles mark the position of prominent elliptical galaxies.    
      The small panels A, B, D and E show zoomed-in regions of a filtered version of the sum of JWST's SW filters. Yellow circles mark the position of automatically detected point sources in the filtered images.  The white contours show the masked regions where no point source detection is performed except for the very central region (panel D) where identification of point sources is done visually inside the 1" radius masked region. The mask of the full region containing bright and/or extended sources as well as artifacts is shown in panel C (bottom right).
         }
         \label{Fig1}
\end{figure*}

The paper is organized as follows.
Sect.~\ref{sec_data} describes briefly the JWST data used in this work. 
In Sect.~\ref{sec_GC} we focus our attention on the GCs describing the steps followed to derive the GC catalog. 
We compare the lens model and GC distribution in Sect.~\ref{sec_lens}, including also a simple recipe to infer the lensing mass from the distribution of GCs.  Section~\ref{sec_contamination} estimates the level of contamination from background galaxies that is needed in order to correct the one-dimensional density profile. We discuss the one dimensional profiles in  Sect.~\ref{sec_radial}. Finally, Sect.~\ref{sect_conclusions} summarizes our conclusions.
We adopted a standard flat cosmological model with $\Omega_M=0.3$ and $h=0.7$. At the redshift of the lens ($z=0.348$), and for this cosmological model, one arcsecond corresponds to 4.921~kpc. 

{\section{JWST data}
\label{sec_data}
AS1063 was observed in 2024 by JWST as part of the Cycle2 GLIMPSE program (PI Hakim Atek, Program ID G0-3293). An overview of the GLIMPSE program and details of the data are given in \cite{Atek2025}. Typical $5\sigma$ depth for a point source ($0\farcs2$ aperture diameter) is  30.9 mag across 9 filters: F090W, F115W, F150W, F200W, F277W, F356W, F410M, F444W, F480M totaling 120 hours of exposure time.

This JWST data were reduced in a similar fashion as in the EPOCHS series of papers \citep[e.g.][]{conselice2024, Adams_2025}. We use a modified version of the official JWST pipeline version 1.8.2, using Calibration Reference Data System (CRDS) pmap1364 to a pixel scale of 0\farcs03 pix$^{-1}$. To better preserve the light from the foreground cluster and prevent over-subtraction due to the bright central galaxy we do not include 1/f subtraction, and we employ a custom 2d background subtraction between stages 2 and 3 (Goolsby et al. in preparation) . We further include the subtraction of custom wisp templates \citep{Adams_2025}, and the images are aligned to GAIA DR3 during stage 3 \citep{GAIADR3}.

\section{Globular clusters in AS1063}
\label{sec_GC}

\begin{figure} 
  \includegraphics[width=9cm]{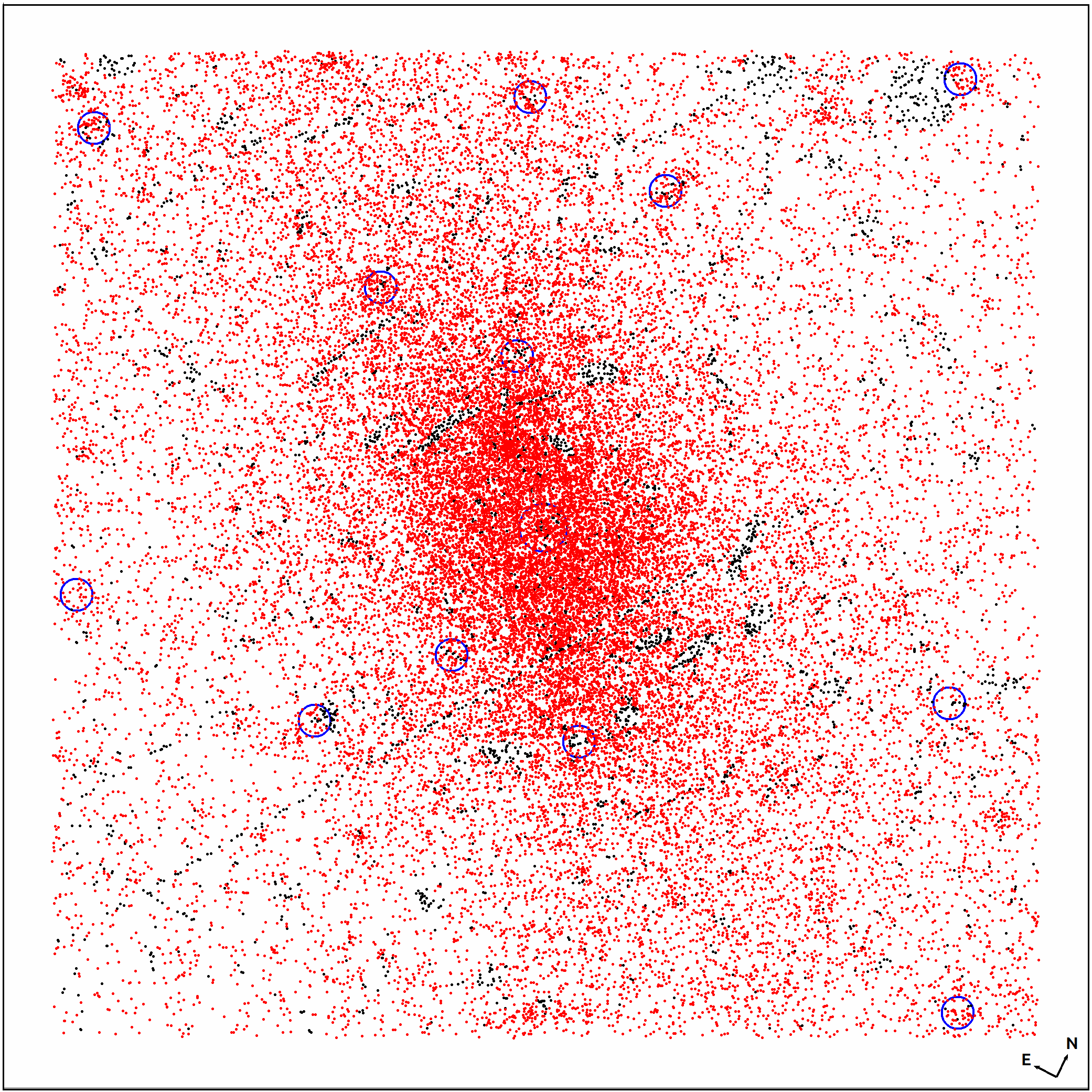}   
      \caption{Map of GC distribution obtained from the data (red points, 28026) and inpainted (black points, 3501) in the masked regions. Some artifacts (diffraction spikes) and large lensed galaxies (giant arcs) are obvious in the distribution of inpainted points. Big blue circles mark the position of prominent members (white circles in Fig.~1).
         }
         \label{Fig_GC_distribution}
\end{figure}

We used the short wavelength filters,  F090W, F115W, F150W, and F200W, to detect GCs since they have the best spatial resolution and sensitivity to unresolved sources in JWST. We applied two steps of high pass filters to remove signal above some spatial scale and boost the signal-to-noise ratio for point sources. In particular, we used the standard difference of two Gaussian smoothed versions of the same image, also known as À Trous wavelet. This is equivalent to a high pass filter in Fourier space and very similar to the popular Mexican hat wavelet. 
Before applying the filter, we masked bright sources (by setting a threshold in the filtered images) and artifacts that otherwise would result in spurious point sources in our final map. Among the artifacts, we masked bright diffraction features around stars and other bright sources (such as AGNs). We also masked other bright extended galaxies (foreground, member, and background galaxies) and all known and candidate lensed galaxies from paper-I, in particular thin giant arcs that, similarly to the case of the diffraction spikes, would result in spurious point sources along the arc after the filtering process. These arcs also contain many unresolved sources that could be confused with GCs, so they need to be masked. We also masked the cores of bright member galaxies and large spirals in the cluster, which can also result in point sources after the filter is applied. Finally, we masked the edges of the image (rectangular regions of width $\approx 6".6$), since these have significantly higher noise and result in more spurious detections. \\

A small level of contamination is still expected after the masking, but this should come mostly from compact dwarf galaxies in the cluster, which nevertheless serve also as tracers of the potential and often contain GCs as well. Also, compact dwarf galaxies are much less common than diffuse ones, with the latter being effectively suppressed by the filtering process. 
More importantly, small high-z galaxies are a true contamination to our sample, but these should have a rather uniform distribution across the field of view. We discuss the level of contamination in section~\ref{sec_contamination} where we use the data in the second module to assess the level of contamination. The final mask in the main module is shown in the bottom right panel (C) of Fig.~\ref{Fig1}. 
After masking, we built the high-pass filtered image. In the first step, we started with a filter $G_{0.5}-G_5$ where $G_{0.5}$ is a Gaussian with FWHM=0.5 pixels and $G_5$ is a Gaussian with FWHM=5 pix. The first scale (0.5 pix) reduces slightly the instrumental noise while the second scale (5 pix) removes structures larger than 5 pixels. At this point, the filtered image shows large positive peaks at the position of luminous sources. If these were not already included in our mask, we added these regions to the mask and remove these luminous sources. 
The resulting image has most of the large scale structure removed, or masked, but still shows diffuse emission around bright sources with a large gradient in the light, specially around the BCG. 
To reduce this diffuse emission even further we performed a second high pass filter similar to the previous one but with scales 1 and 3 pixels, that is, $G_1-G_3$. As before, $G_1$ reduces the contribution from noisy pixels while $G_3$ removes the unwanted diffuse emission. The filter $G_1-G_3$ sharpens the point sources facilitating their detection. \\

The final step is to detect the point sources in the filtered detection map. This map is built after adding the double-filtered images of F090W, F115W, F150W, and F200W. Equal weights are assumed for each band, which corresponds to an agnostic prior on the spectra of the GCs. 
We started by the brightest pixel, recorded it, and masked the area around that pixel with a disk of radius 3 pixels. Then we found the new maximum and repeated the process until a predetermined threshold was reached. We defined the threshold as the point where detections start to show a regular pattern associated with the CCDs, as this is a clear sign that we are reaching the noise detection level. Even though we stopped the detection at this point, the final catalog of GCs contains only objects detected at twice this threshold to reduce the number of spurious detections, but among the discarded sources there are definitely more, but faint, GCs. \\

The detected GCs are shown in Fig.~\ref{Fig_GC_distribution} as red dots. A clear concentration is found around the BCG with the GC distribution stretching in the same NE direction as the lens model shown in Fig.~\ref{Fig1}. 
A direct comparison of the surface number density of GCs, $\Sigma_{\rm GC}$ vs the local average of the lens model convergence,  $<\kappa>$, reveals a strong correlation, as shown in Fig.~\ref{Fig_GCNumberDensity_vs_Kappa}.

To derive this result, we first computed the surface density of GCs, ${\rm n}_{\rm GC}$, and average convergence, $<\kappa>$, in boxes of $3"\times3"$. Then we selected areas with similar ${\rm n}_{\rm GC}$ and computed the mean and dispersion of  $<\kappa>$ values in the same regions. We find the scaling law $<\kappa> = 0.4 {\rm n}_{\rm GC}^{0.7}$ fits well the correlation. From this simple relation we see that regions in the lens plane with GC densities ${\rm n}_{\rm GC}>4$ arcsec$^{-2}$ should be supercritical ($\kappa>1$) and contain multiply lensed galaxies.  The relation seems to break at very high values of ${\rm n}_{\rm GC}$ but this is due to the baryonic contribution to the convergence that increases $\kappa$ near the center of halos while the number density of GCs plateaus at the center of halos. In the opposite regime where ${\rm n}_{\rm GC}$ is low, the correlation seems to hold. This is an important result which can be exploited to fill the gap of incomplete lens models with few lensing constraints in the outskirts of the cluster. The relation between ${\rm n}_{\rm GC}$ and $\kappa$ may be also used in combination with strong lensing, for instance in two merging clusters, A \& B, where one of the clusters is rich, A, in lensing constraints but the other one, B, is not. The well constrained cluster A can be used to calibrate the  ${\rm n}_{\rm GC}$--$\kappa$ relation which can later be used to infer the mass distribution of the poorly constrained B cluster, from its system of GCs and the above relation. This approach should be tested in merging clusters with lensing constraints in both subclusters, although early results on the merging cluster MACS0416 suggest that this approach should be feasible \citep[see Fig.~8 in][]{Diego2024a}

\begin{figure} 
  \includegraphics[width=9cm]{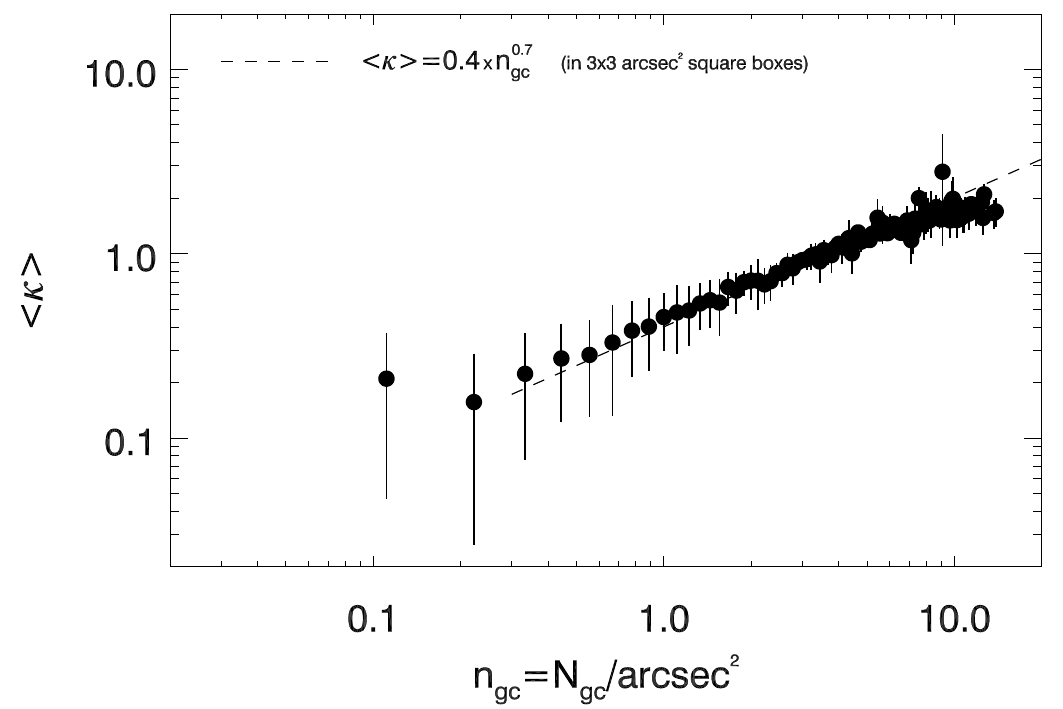}   
      \caption{Number density of GCs vs mean convergence computed in boxes of $3"\times3"$. The dashed line corresponds to $y=0.4x^{0.7}$ and fits reasonably well the observed correlation.
         }
         \label{Fig_GCNumberDensity_vs_Kappa}
\end{figure}
The final number of GCs (red points) in Fig.~\ref{Fig_GC_distribution} is 28026 but this does not account for the number of missing GCs in the masked regions, which we estimated it to be $\approx 3500$. We obtained this number of masked objects and their spatial distribution by first computing the density of GCs in every pixel in the image, $\rho_{\rm gc}(x,y)$, and then doing a Monte Carlo to simulate the objects contained in the masked region.

To compute the continuous 2D surface density of GCs, $\rho_{\rm gc}$, we convolved the discrete distribution by a function that depends on the inverse of the distance to a GC,
\begin{equation}
\rho_{\rm gc}(x,y) = \int F(x,y)W(x-x',y-y')dx'dy'\, ,
\label{eq_rho}
\end{equation}
where $F(x,y)=1$ at the position of a GC and 0 otherwise and $W$ is the convolution kernel,
\begin{equation}
W(x-x',y-y') = \frac{1}{(S_L + d)^{\alpha}}\, ,
\label{eq_kernel}
\end{equation}
with $d$ the distance between positions $(x,y)$ and $(x',y')$ and $S_L$ is a constant representing the softening length, needed to avoid the convolution kernel to diverge for $d=0$. This approach is similar to Sepherd's algorithm where $\alpha=2$ is usually (but not necessarily) adopted. In our case, we are interested in a value of $\alpha$ that results in $\rho_{\rm gc}(x,y)$ reproducing the mass profile from the lens model, so $\rho_{\rm gc}(x,y)$ can be also used as a proxy for the lensing mass in future clusters with poor or no lensing constraints. We find that a value of $\alpha=1.6$ results in a good fit to the lens model profile (see section~\ref{sec_radial}). 
The value of the softening length is not as critical and values around $0".5<S_L<1"$ produce satisfactory results. For this work we used $S_L=0".9$. \\

The convolution is efficiently computed in Fourier space. A first estimation of the density can be obtained even when there are holes in the distribution of GCs, as for instance in the masked regions. The relatively slow decline of the kernel with distance, $d^{-\alpha}$, guarantees that even in the center of large masked regions the density is not much smaller than in the edges. The density in these regions is then a relatively good approximation to the expected density of GCs in the masked regions. We can use this first estimation of the density to fill the holes, or inpaint the masked region, through a Monte Carlo where we randomly place GCs in the masked regions according to the pre-estimated density. The final result is shown in Fig.~\ref{Fig_GC_distribution} with red points showing the original distribution of GCs and black points showing the inpainted GCs in the masked region. The process can be iterated several times until the density converges, but for our particular case, one simple iteration produces satisfactory enough results. The edge portion of the mask is not inpainted since it is noisy and contains too many spurious sources.

\section{Lens model vs GC distribution}
\label{sec_lens}
Following \cite{Harris2017}, the total mass in GCs should be $2.9\times10^{-5}$ times the total mass of the galaxy cluster. This relation is applied at the  virial radius and for the halo virial mass. Neither the lens model nor the mapping of the GCs extend to the virial radius (without gaps in the data such as the gap between modules 1 and 2 in JWST data), so we take this relation as an approximation. Instead we consider a radius of 60" which is the maximum distance from the center at which we can detect GCs in the main module. The total mass contained in the lens model up to 60" is $4.23\times10^{14}\, \Msun$, so the total mass in GCs should be $2.9\times10^{-5}$ times smaller, or $8.47\times10^{9}\, \Msun$. Within the same radius, we find $27895$ GCs (including the inpainted ones), so the average mass per GC is $\mathcal{M}=3.03\times10^{5}\, \Msun$, or in log scale, ${\rm log}_{10}(\mathcal{M})=5.48$. This is just a little above the typical mass of GCs in the Milky Way, where \cite{Baumgardt2018} finds a mean log mass of ${\rm log}_{10}(\mathcal{M})=5.2$ and dispersion $\sigma_{{\rm log}(\mathcal{M})} = 0.5$, and comparable to expectations from N-body simulations \citep{Pfeffer2018}. \\

\begin{figure} 
  \includegraphics[width=9cm]{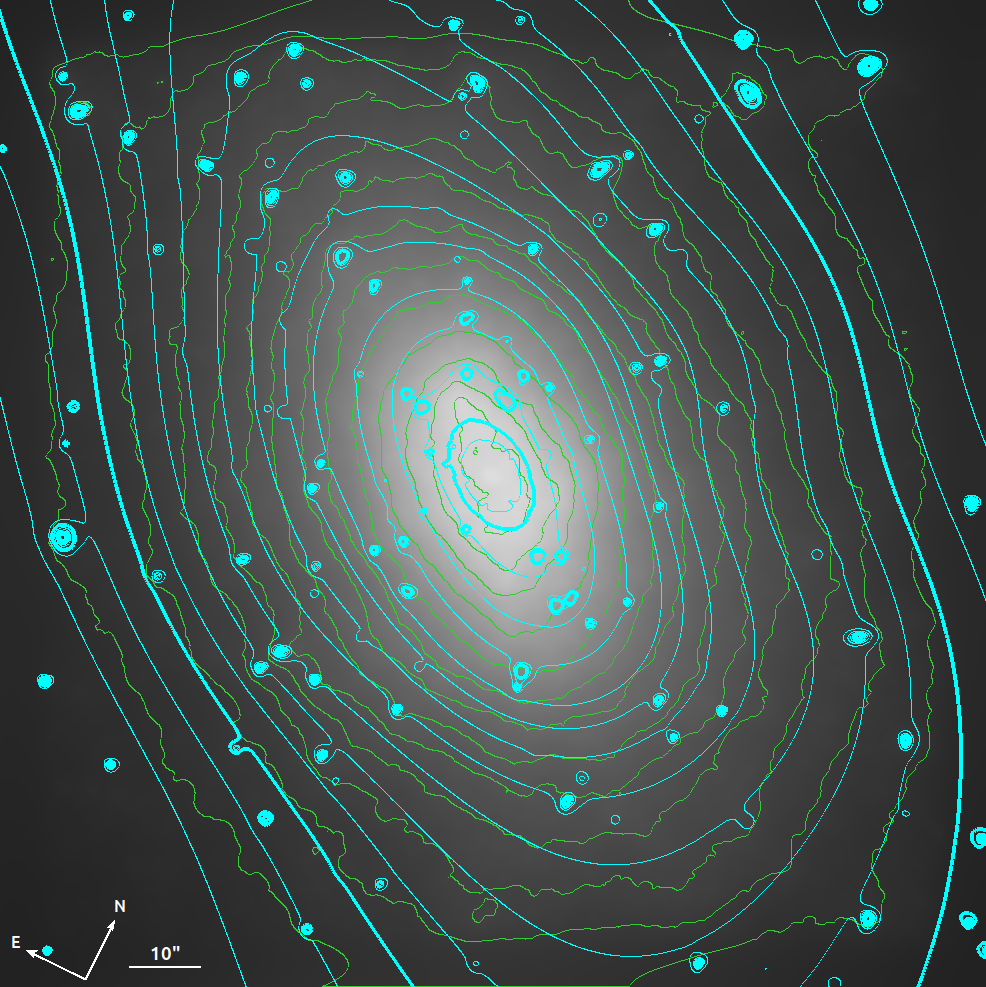}   
      \caption{Map of GC density after inpainting the masked regions (gray). The green contours are for the GC density map (gray scale image) while the cyan contours are for the lens model convergence map from paper-I. The two thicker cyan contours correspond to $\kappa=0.3$ and $\kappa=1.8$.
         }
         \label{Fig_kappa_vs_rho_contours}
\end{figure}

Using the observed distribution of GCs (red points in Fig.~\ref{Fig_GC_distribution}) plus the inpainted GCs (black points in the same figure), we inject these distributions into Eq.~\ref{eq_rho} and recompute the 2D density map of GCs in the entire field of view. 

The result is shown in Fig.~\ref{Fig_kappa_vs_rho_contours} in gray color. The contours of the GC density are shown in green and we compare them with the contours of the convergence, $\kappa$, from the lens model (paper-I) in cyan. The agreement between contours is remarkable, confirming that the distribution of GCs can be used as an alternative way of inferring the distribution of dark matter, even in the absence of strong lensing constraints. This technique may be particularly useful for low redshift clusters in which the low redshift results in subcritical surface mass densities and hence no strong lensing, while on the other hand the low redshift is a great ally for the detection of more GCs in the galaxy cluster. 

An alternative, more quantitative way, of showing the tight relation between $\rho_{\rm gc}$ and $\kappa$ is shown in Fig.~\ref{Fig_kappa_vs_rho} where we plot the pixel by pixel correlation between $\rho_{\rm gc}(x,y)$ and $\kappa(x,y)$. For this comparison, we renormalize $\rho_{\rm gc}$ by the usual constant, $\beta$, which minimizes the variance of the difference $\Delta=\kappa-\beta\times\rho_{\rm gc}$ and is simply given by:
\begin{equation}
\beta=\frac{\int\kappa(x,y)\rho_{\rm gc}(x,y)dxdy}{\int\rho_{\rm gc}^2(x,y)dxdy}
\end{equation}
The correspondence between $\rho_{\rm gc}$ and $\kappa$ is very good for values of the convergence above ${\rm log}_{10}(\kappa)\approx-0.5$ or $\kappa \approx 0.3$. For smaller values of $\kappa$ the discrepancy is partially due to the shallower slope of the assumed kernel ($\alpha=1.6$ in Eq.~\ref{eq_kernel}) which extrapolates the density at large radius (where GC are not detected or masked by the edge portion of the mask), while DM is known to fall faster ($\alpha\approx2$ or NFW-like). Also, at large radii the relative contribution from background small galaxies, which are misidentified as GC by our filtering algorithm become more relevant. We address the possible level of contamination from foreground objects in the next section.

\begin{figure} 
  \includegraphics[width=9cm]{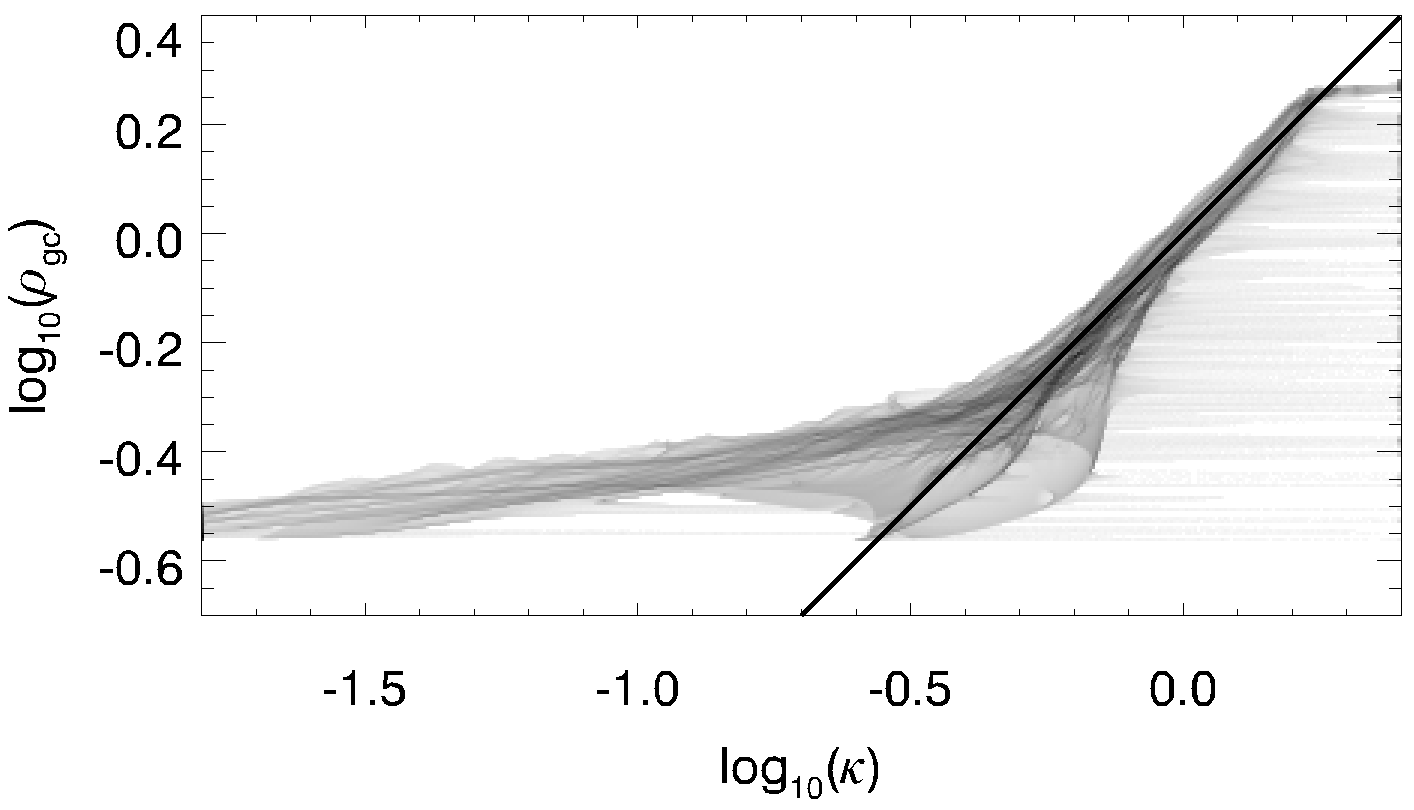}   
      \caption{Scatter plot of convergence vs GC density maps. The solid line is the $x=y$ model. The GC density, $\rho_{\rm gc}$  has been re-scaled by the factor $\beta$ that minimizes the variance of the difference map $r=\kappa-\beta\times\rho_{\rm gc}$. 
         }
         \label{Fig_kappa_vs_rho}
\end{figure}

\section{Contamination level}\label{sec_contamination}
We repeat the same process described in section~\ref{sec_GC} but in the second module of the GLIMPSE data. This second module allows us to detect GCs from $1.'8$ (0.53 Mpc) to $4'$ (1.18 Mpc) from the cluster center. Several bright stars fall in this module that increases the size of the masked region. On the other hand, fewer cluster members and no giant arcs are found in this module. The number density of high redshift and compact galaxies (that can contaminate our sample) is supposed to be the same as in the main module as long as  the luminosity function of high-redshift galaxies is close to a power law with slope -2. This would compensate for the magnification bias resulting in a uniform distribution of lensed galaxies.  
The resulting mask and detected GCs candidates are shown in Figure~\ref{Fig_GCmask_Paralell}. In total we detect 5327 GCs and estimate that 804 should have been detected in the masked region. The average number density in the second module is ${\rm n}_{\rm GC}=1.67\times10^{-2}$ kpc$^{-2}$. The spatial distribution shows a mild gradient in the north-south direction with a slight increase in number density in the direction of the cluster center.

With the detection of GCs in the second module we estimate the level of contamination from compact background galaxies as the surface number density of GCs at the largest radii (0.015 GC per kpc$^2$, see next section). This represents a conservative upper limit to the contamination, since some of these distant detections are real GCs in the outskirts of the cluster.

\section{The radial distribution of GCs}
\label{sec_radial}

\begin{figure} 
  \includegraphics[width=\linewidth]{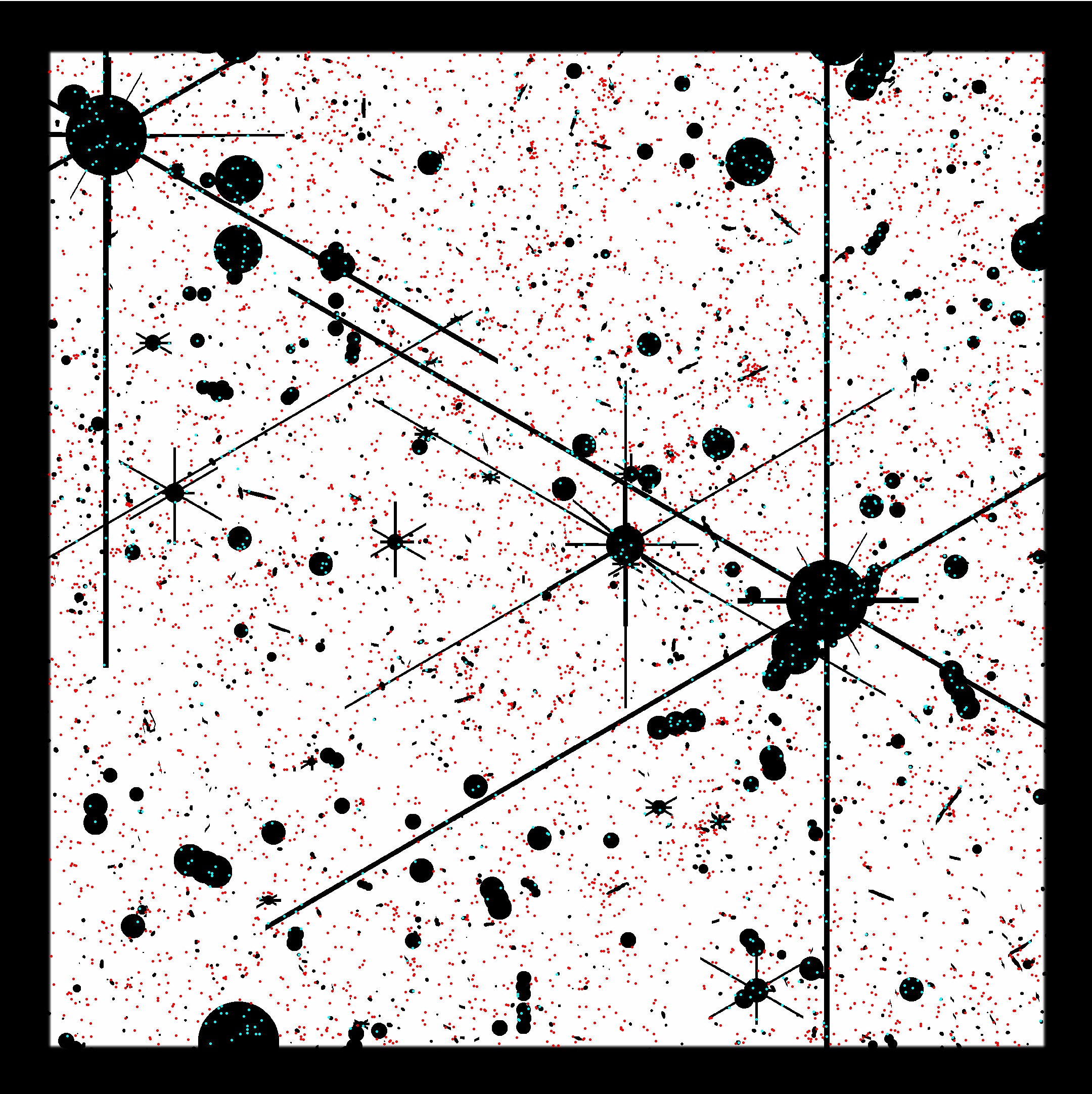} 
      \caption{Mask (black) and detected GC (red dots, 5327) in the second module. The cyan dots are the inpainted GCs in the masked region (804).  The orientation is the same as in Fig.~\ref{Fig1}. 
         }
         \label{Fig_GCmask_Paralell}
\end{figure}

We show the one dimensional profiles of the GC distribution and lensing mass in Fig.~\ref{Fig_Profiles_V3}. 
The lensing mass profile derived in paper-I (convergence at $z_s=3$) is shown as a dark blue line while the profile of GCs is shown as a thick black solid line. For easier comparison with the lensing convergence, we have re-scaled the number density of GCs by a factor 4. The inner 1" is masked during the process of detecting clusters but we perform a  visual inspection in this region of the filtered maps and detect GCs by eye in this small region (see Fig.~\ref{Fig1}). This central 1" shows a small but noticeable deficit of GCs when compared with neighboring regions. This could be due to the higher photon noise at the center of the BCG which makes detection of GCs more challenging.  Although potentially interesting, this apparent deficit in the number of GCs in the central $1"$ is not reliable so we ignore distances smaller than $1"$ in our analysis.  We observe the number density of GCs remains constant between $3"$ and $13"$. This is similar to the trend observed in the lensing mass. The number density of GCs starts to decline above $r\gtrsim13"$. The decline in the lensing mass, or convergence, starts at larger distances. Between $13"<r<60"$ the GC number density falls approximately as the expected $r^{-2}$ found in simulations \citep{ReinaCampos2023}. \\

At even larger radii, we can derive the profile from the second module of JWST-GLIMPSE that extends the field of view to $\approx4'$ from the center and close to the virial radius of the cluster. We repeat the same process as in the main module  and derive the profile of the number density of GCs. This is shown in the same figure as a thick solid black line between radii $100" \lesssim r \lesssim 250"$.  We can use the profile at the largest radius as an upper limit for the level of contamination from small compact background objects, which also enter our GC catalog. We estimate the level of contamination as 0.015 GC per kpc$^2$ (or 14.7\% the average number density of GCs in the inner 60" radius region), shown as a horizontal black dotted line in the figure. This value is above, but still close to, the background contamination level estimated in the outskirts of the Coma cluster from shallower HST-ACS imaging in \cite{Peng2011}. The corrected GC profiles obtained after subtracting this level of contamination are shown as a thin black solid line. 
At radii $r>50"$, the lensing mass from our model in paper-I falls faster than $r^{-2}$. This is due to a known bias in our lensing reconstruction beyond the maximum radius constrained by lensing, but if GC are true tracers of dark matter, the corrected GC profile shown in Fig.~\ref{Fig_Profiles_V3} suggests that around $r=100"$ the mass profile should still fall close to $r^{-2}$. 
For comparison we also show the profile of the light (including the intracluster light) as observed in the F444W filter that has a similar decline with radius as the GC distribution. However, being much fainter than individual GCs, the light profile can only be detected at shorter distances from the center of the cluster \citep[see also][]{Montes2019}, while GCs can be detected all the way (and beyond) the virial radius. 

\begin{figure} 
  \includegraphics[width=\linewidth]{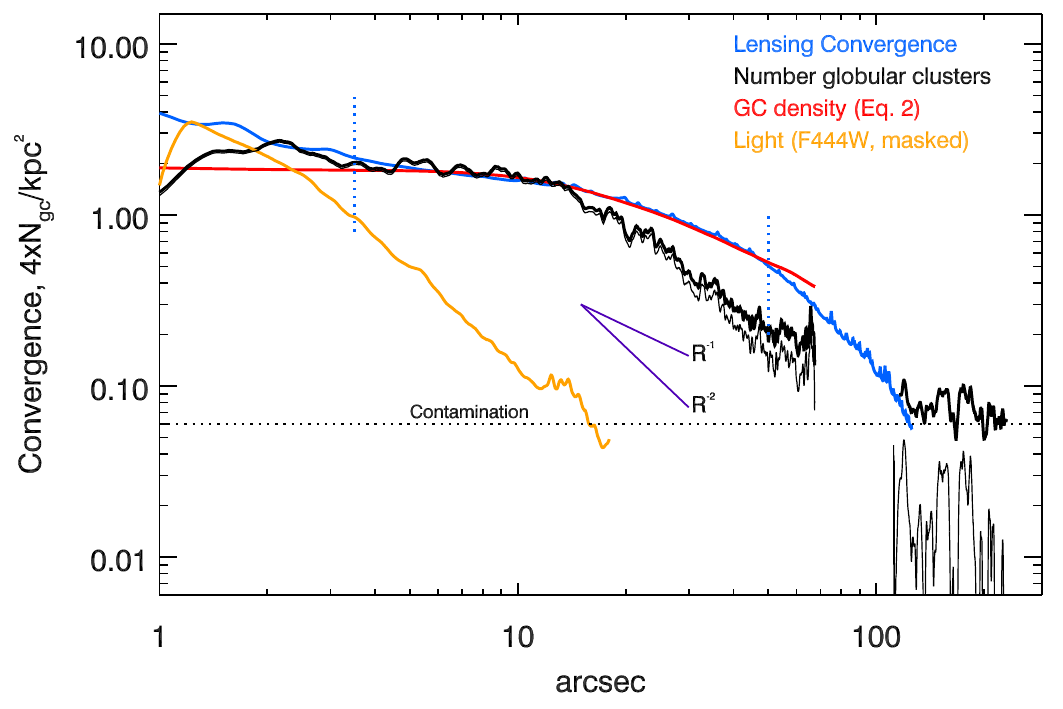} 
      \caption{Profiles of AS1063 centered on the BCG. The projected mass from the lens model is plotted as dark blue (in units of $\Sigma/\Sigma_{\rm crit}$) with the two blue vertical dotted lines marking the range of distances with lensing constraints. The light profile (including the intracluster light) in the filter F444W is shown as an orange line. The map of the light is masked and only the unamasked pixels are used for the profile. The masked BCG results in a nil light profile at the very center. The number density of GCs is shown as a thick black line that is re-scaled by a factor four to match the convergence profile for easier comparison. The gap in the profile corresponds to the gap between the two modules where no GCs are detected. The GC number density profile at large radii ($100" \lesssim r \lesssim 250"$), is computed in the second module. After subtracting the estimated level of contamination from background objects (horizontal dotted line), the resulting corrected profile is shown in black as a thin solid line. 
         }
         \label{Fig_Profiles_V3}
\end{figure}
The light profile (or baryons, orange line) determines the shape of the lensing model (blue line) in the inner $r<2"$, but beyond this radius the projected mass profile is dominated by the contribution from DM. From this plot, it is evident that the mismatch between the intracluster light (core or few kpc) and projected mass in the lens model (large core). Meanwhile, the GC distribution, although with a somewhat smaller core than the lensing convergence profile, resembles much better the distribution of dark matter and shows also a large core. 
The similarity between the GC and DM profiles is better highlighted when we compare the profile of the continuous 2D distribution of GCs, or $\rho_{\rm gc}$ (given by Eq.~\ref{eq_rho}) with the lens model. This is shown as a red curve in Fig.~\ref{Fig_Profiles_V3}. The agreement with the lens model is excellent in the region constrained by lensing. Above $r\approx 50"$ the red curve ($\rho_{\rm GC}$) is above the blue one ($\kappa$) due mostly to the biased low lensing solution beyond the constrained region and the shallower decline in the smoothing kernel of $\rho_{\rm GC}$ at radii where we do not have GCs. 

\section{Conclusions}\label{sect_conclusions}
We take advantage of the exceptionally deep images taken as part of JWST's GLIMPSE program in AS1063, to detect the faint population of GCs in this cluster. We find tens of thousands of GCs in the central module and thousands in the second module that extend the coverage up to distances close to the virial radius. We find a good correlation between the spatial distribution of GCs and lensing mass. We define a smoothing kernel that can transform the discrete distribution of GCs into a continuous distribution that resembles the lensing convergence. Similar transformations can be used in other clusters in which lensing constraints are not available (for instance in low redshift subcritical clusters) in order to get a proxy for the mass distribution. We compare pixel-by-pixel the smoothed GC distribution with the lensing convergence and find a good correspondence.
We also detect GCs in the second module and set the upper limit of the contamination from background objects as the density of GCs at the largest radius. After correcting for this contamination, the derived GC profile falls approximately as $r^{-2}$ up to the largest radius probed by our analysis. \\

GCs offer an alternative proxy for the spatial distribution of lensing mass in clusters, a fact that can be exploited in future analysis of galaxy clusters that show, or do not show, strong lensing features.

\begin{acknowledgements}
J.M.D. acknowledges the support of projects PID2022-138896NB-C51 (MCIU/AEI/MINECO/FEDER, UE) Ministerio de Ciencia, Investigaci\'on y Universidades and SA101P24. 

This work is based on observations made with the NASA/ESA/CSA \textit{James Webb} Space Telescope. The data were obtained from the Mikulski Archive for Space Telescopes at the Space Telescope Science Institute, which is operated by the Association of Universities for Research in Astronomy, Inc., under NASA contract NAS 5-03127 for JWST. 
These observations are associated with JWST program \#3293. 

\end{acknowledgements}

\bibliographystyle{aa} 
\bibliography{MyBiblio} 

\end{document}